%% file: msr.tex
\documentclass[10pt,conference]{IEEEtran}
\IEEEoverridecommandlockouts
\usepackage{cite}
\usepackage{amsmath,amsfonts,amsthm}
\usepackage{algorithmic}
\usepackage{graphicx}
\usepackage{textcomp}
\usepackage{xcolor}
\usepackage{pifont}
\usepackage{enumitem}
\usepackage{balance}
\usepackage{booktabs}
\usepackage{arydshln}
\usepackage{threeparttable}
\usepackage{makecell}
\usepackage{colortbl}
\usepackage{xspace}
\usepackage[many]{tcolorbox}
\usepackage{hyperref}
\usepackage{tikz}
\usepackage{gensymb}
\usepackage{multirow}
\usepackage{subcaption}

\def\BibTeX{{\rm B\kern-.05em{\sc i\kern-.025em b}\kern-.08em
    T\kern-.1667em\lower.7ex\hbox{E}\kern-.125emX}}

\newcommand{\xrappdataset}{\textsc{XRZoo}\xspace}
\newcommand{\xrappdatasetnumber}{12,528\xspace}
\newcommand{\xrzooappstorenumber}{nine\xspace}

\makeatletter
\def\adl@drawiv#1#2#3{%
	\hskip.5\tabcolsep
	\xleaders#3{#2.5\@tempdimb #1{1}#2.5\@tempdimb}%
	#2\z@ plus1fil minus1fil\relax
	\hskip.5\tabcolsep}
\newcommand{\cdashlinelr}[1]{%
	\noalign{\vskip\aboverulesep
		\global\let\@dashdrawstore\adl@draw
		\global\let\adl@draw\adl@drawiv}
	\cdashline{#1}
	\noalign{\global\let\adl@draw\@dashdrawstore
		\vskip\belowrulesep}}
\makeatother

\definecolor{applegreen}{rgb}{0.55, 0.80, 0.4}

\let\diff\undefined %

\ifdefined\diff

\else

\fi

\definecolor{mygray}{gray}{.9}

\begin{document}

\title{\xrappdataset: A Large-Scale and Versatile Dataset of Extended Reality (XR) Applications}
\author{\IEEEauthorblockN{Shuqing Li}
\IEEEauthorblockA{
\textit{The Chinese University of Hong Kong}\\
sqli21@cse.cuhk.edu.hk}
\and
\IEEEauthorblockN{Chenran Zhang, Cuiyun Gao}
\IEEEauthorblockA{
\textit{Harbin Institute of Technolgy}\\
\{220110514@stu, gaocuiyun@hit\}.edu.cn}
\and
\IEEEauthorblockN{Michael R. Lyu}
\IEEEauthorblockA{
\textit{The Chinese University of Hong Kong}\\
lyu@cse.cuhk.edu.hk}
}

\maketitle

\input{files/abstract}


\input{files/introduction}
\input{files/approach}
\input{files/free_dataset}

\input{files/discussion}
\input{files/conclusion}

\balance
\bibliographystyle{IEEEtran}
\bibliography{msr}

\end{document}

%% file: files/abstract.tex
\begin{abstract}
The rapid advancement of Extended Reality (XR, encompassing AR, MR, and VR) and spatial computing technologies forms a foundational layer for the emerging Metaverse, enabling innovative applications across healthcare, education, manufacturing, and entertainment.
However, research in this area is often limited by the lack of large, representative, and high-quality application datasets that can support empirical studies and the development of new approaches benefiting XR software processes. 
In this paper, we introduce \xrappdataset, a comprehensive and curated dataset of XR applications designed to bridge this gap. 
\xrappdataset contains \xrappdatasetnumber free XR applications, spanning \xrzooappstorenumber app stores, across all XR techniques (i.e., AR, MR, and VR) and use cases, with detailed metadata on key aspects such as application descriptions, application categories, release dates, user review numbers, and hardware specifications, etc. 
By making \xrappdataset publicly available, we aim to foster reproducible XR software engineering and security research, enable cross-disciplinary investigations, and also support the development of advanced XR systems by providing examples to developers. 
Our dataset serves as a valuable resource for researchers and practitioners interested in improving the scalability, usability, and effectiveness of XR applications.
\xrappdataset is publicly released and actively maintained, which can be accessed via \href{https://sites.google.com/view/xrzoo}{\color{blue}{https://sites.google.com/view/xrzoo}}.
\end{abstract}

%% file: files/introduction.tex
\section{Introduction}

Extended Reality (XR) applications encompass Virtual Reality (VR), Augmented Reality (AR), and Mixed Reality (MR) applications, relying on spatial computing~\cite{bhowmik2024virtual} to deliver immersive and interactive user experiences~\cite{DBLP:journals/csur/AudaGFMS24}. 
They form a critical foundation for the emerging Metaverse~\cite{dionisio20133d}, a convergence of physical and virtual spaces facilitated by spatial computing.
Such applications leverage sophisticated hardware devices such as head-mounted displays (HMDs), motion sensors, and real-time environmental mapping to enable users to interact with digital content in physical and virtual space.
The three reality technologies differ in their degree of interaction with the physical world: VR fully immerses users in a simulated environment, AR overlays digital information onto the physical world while maintaining user presence in reality, and MR integrates real and virtual elements in real-time, enabling dynamic interaction between the two.

Advancements in XR applications help drive the surge in global market size, with the size reaching \$54.58 billion in 2024 and being projected to reach \$100.77 billion by 2026~\cite{report:global-augmented-virtual-reality-market-size-msr}.
These applications have revolutionized numerous domains, including healthcare, education, manufacturing, and entertainment~\cite{ziker2021cross}.
For instance, in healthcare, XR applications enable surgical simulations and remote diagnostics, while in education, immersive environments enhance interactive learning experiences. Similarly, in manufacturing, XR has facilitated advanced training simulations and real-time collaboration across geographies.

The increasing importance of XR applications makes it urgent to conduct software engineering (SE) and security research, facilitating the reliability and efficiency of the whole software process from design, implementation, testing, and maintenance~\cite{li2023towards, paper:issre-webxr-empirical, nair2024berkeley, wen2024vr}.
Many researchers have been drawn to address XR software challenges across diverse domains, such as automated interactive testing~\cite{li2024grounded, qin2024utilizing}, bug detection~\cite{yang2024towards, li2024less}, and static/dynamic analysis (e.g., for privacy auditing or clone detection)~\cite{guo2024empirical, trimananda2022ovrseen}.

Similar to research in conventional software domains, the primary obstacle lies in subject collection.
To the best of our knowledge, there doesn't exist a large-scale XR application dataset in the research community.
The fragmented nature of dispersed XR ecosystems, coupled with diverse development technology stacks and competing hardware platforms, creates platform-specific restrictions on accessing XR applications and their associated data. This fragmentation severely limits the ability to collect comprehensive datasets, resulting in such lack of accessible, high-quality datasets that span the broad spectrum of XR platforms and application categories.
As a result, researchers often resort to reusing limited datasets (has risks of being outdated) or collecting small, non-representative samples. Such approaches can lead to biased analyses and experiments that are difficult to reproduce, thereby limiting the generalizability and impact of research in this domain.

The datasets with great usability in other software domains,
including mobile apps~\cite{paper:androzoo,paper:androzooopen,paper:android-dataset-3}, video games~\cite{paper:playmydata}, and microservices~\cite{paper:conf/msr/dAragonaBLSAABBKLTWNQRAMCT24}, etc., inspires us to fill the gap.
In this paper, we introduce \xrappdataset, the first large-scale curated
dataset that comprehensively covers XR applications across different application categories, ecosystems, technology stacks, and hardware platforms. 
To facilitate reproducibility and ease of access, we present a well-documented methodology for collecting, processing, and curating the dataset. 
The data collection faces significant challenges in ensuring data quality, completeness, and accuracy due to technical and platform-specific constraints. These include the diversity of XR platforms, which require sophisticated filtering algorithms to identify XR applications, especially on general-purpose stores that lack explicit categorization. Data crawling is further complicated by limited API support and access restrictions, necessitating custom distributed crawling systems to bypass rate limits. Metadata aggregation across platforms is challenging due to inconsistent and variable data formats, requiring additional effort to synthesize and normalize information.
We address these challenges by self-implementing specialized crawlers to collect applications from diverse XR marketplaces, ensuring comprehensive coverage of available apps while respecting platform-specific constraints.

The final dataset includes \xrappdatasetnumber free applications from \xrzooappstorenumber leading XR marketplaces, including Meta Horizon Store \& App Lab, Vision Pro App Store, VIVEPORT, Microsoft Store, SideQuest, Steam, Google Play, and the iOS App Store. 
The resulting dataset is diverse in terms of application categories (e.g., productivity tools, educational software, and games), XR device types (standalone, PC-assisted, and mobile devices), and user engagement metrics.
\xrappdataset can support a wide range of SE research, from interactive testing, bug detection, static/dynamic analysis, performance analysis, and usability studies to automatic classification of XR applications.

%% file: files/approach.tex
\section{Data Collection Architecture}

\begin{table}[t!]
    \centering
    \caption{XR devices in ecosystem supported by \xrappdataset}
    \vspace{-0.8em}
    \label{table:xrzoo-devices-support}
    \resizebox{\linewidth}{!}{
    \begin{tabular}{p{2.9cm}p{7cm}}
        \toprule
        \textbf{Supported App Store} & \textbf{Correspondng XR Devices} \\
        \midrule
        \rowcolor{mygray}
        Meta Horizon Store \&  Meta Horizon App Lab & Meta Quest series (Quest 2, Quest Pro, Quest 3, Quest 3S); Oculus Series (Quest, Rift, Rift S, GearVR, Go); and other devices running Meta Horizon OS (e.g., upcoming devices from ASUS and Lenovo). \\
        VIVEPORT & HTC Vive series (VIVE, VIVE Cosmos, VIVE Pro, VIVE Focus 3, VIVE XR Elite), Meta Quest series (via PC Link), Valve Index, and Windows Mixed Reality. \\
        \rowcolor{mygray}
        SideQuest & Primarily supports Meta Quest series and Oculus Quest; and other Android-based standalone VR headsets like Pico Series (Pico Neo 3, Pico 4, Pico 4 Pro). \\
        Vision Pro App Store & Apple Vision Pro. \\
        \rowcolor{mygray}
        Steam & More than 30 VR headsets that support OpenVR, including HTC Vive, Meta Quest Series (via PC Link), Pico Series (via PC Link), Valve Index, and Windows Mixed Reality, etc. \\
        Microsoft Store & HoloLens 2 and Windows Mixed Reality headsets. \\
        Google Play & Android devices supporting ARCore or VR applications. \\
        iOS App Store & iOS devices supporting ARKit or VR applications. \\
        \bottomrule
    \end{tabular}
    }
    \vspace{-1.5em}
\end{table}

\subsection{Data Source}

To build \xrappdataset, we collect from multiple mainstream marketplaces that host XR applications. 
Our goal is to cover a diverse and comprehensive set of XR applications that can be retrieved freely.
Specifically, we target marketplaces across XR-specific app stores, general-purpose game app stores, and mobile phone app stores.
We try our best to go through all mainstream marketplaces and determine \xrzooappstorenumber app stores at last.
The selection criteria are the following:
(1) The marketplace hosts XR applications that cannot be found elsewhere (some app stores are only collections of the original marketplaces).
(2) The marketplace has active users (e.g., have reasonable numbers of application downloads or user reviews).
Table~\ref{table:xrzoo-devices-support} summarizes the app stores crawled for \xrappdataset, along with the corresponding devices in the XR ecosystem that are supported by the dataset.

\subsubsection{XR-Specific App Stores} For XR-specific app stores, which primarily host XR applications, we collect all their available free applications. These app stores are specialized platforms, mainly maintained by XR device manufacturers, to promote the XR software ecosystem. These stores consist of Meta Quest App Store (and Meta Quest App Lab), VIVEPORT, and SideQuest. These are the most popular domain-specific platforms for XR content distribution.

Specifically, Meta Horizon Store is the primary distribution platform for XR devices from Meta, the world's leading XR company.
Meta Horizon App Lab serves as an alternative distribution channel for Quest applications that have not undergone the full curation process required for the main store. It enables developers to distribute experimental or early-access content directly to users.
The Vision Pro App Store is a specialized marketplace designed to host MR applications optimized for immersive spatial computing experiences on the Vision Pro device.
VIVEPORT is a comprehensive VR content platform developed by HTC.  VIVEPORT applications can be accessed by standalone XR devices via a PC connection (e.g., using Meta Quest Link).
SideQuest is a third-party platform facilitating the sideloading of applications onto Meta Quest devices, SideQuest serves as an early access layer for XR content, enabling developers to distribute applications outside the official Meta ecosystem. 

\subsubsection{General-purpose App Stores on Windows} 
Steam, being one of the largest platforms for digital distribution of games, hosts a wide range of VR applications. 
Microsoft's HoloLens utilizes the Microsoft Store as its application distribution platform, providing a range of MR applications tailored for enterprise and developer use. 

\subsubsection{Mobile Phone App Stores} 
Google Play and iOS App Store are prominent mobile application marketplaces that host a wide range of applications, including VR/AR applications that can be accessed on mobile phones. 

\subsection{Collection Challenges and Our Solutions}

The collection of \xrappdataset
presents several challenges that require careful consideration to ensure data quality, completeness, and accuracy. These challenges stem from both technical constraints and platform-specific limitations as follows.

\subsubsection{Platform Diversity and Data Availability}
One of the primary challenges in collecting XR applications is the inherent diversity across the platforms hosting these applications. While the number of XR platforms continues to grow, each platform presents unique challenges in terms of crawling methods, data accessibility, metadata completeness, and application categorization, etc. For instance, VR-specific platforms like Meta Quest and VIVEPORT offer a more direct pathway to identifying XR applications.
Additionally, app stores with broader content, such as Microsoft Store, Steam, and mobile app stores, require precise filtering mechanisms to isolate XR applications from a large pool of non-XR games. This necessitated the development of sophisticated filtering algorithms that rely on both (1) explicit tags (e.g., ``supported devices'', if any), and (2) deeper analysis like XR-related library detection on the binary files (detailed below).
Following this strategy, we filter XR applications from Microsoft Store (with ``supported devices'' as Apple Vision Pro) and Steam (with ``supported devices'' as ``VR Support'' or ``VR Only'').

\subsubsection{Complexity of XR Application Identification on Mobile Phone Marketplaces}
The detection of XR applications on mobile phone platforms, i.e., Google Play and iOS App Store, posed additional challenges due to the lack of explicit categorization or tagging for VR/AR applications. Unlike XR-specific platforms, where applications are clearly marked for XR usage, mobile stores often classify XR applications within general-purpose categories, such as ``Utilities'' or ``Games'', without any distinct indication of their XR capabilities. For Android applications, we utilized a widely-used library detection tool (LibScout~\cite{paper:libscout}) to identify those leveraging any Android VR SDKs, However, this approach is not directly applicable to iOS, which is hard to perform library identification. For iOS, we employ a heuristic approach based on matching app titles and developer information from the identified Android applications. 

\subsubsection{Insufficient API Support and Access Restrictions}
Only a few marketplaces provide API support. 
For most of the app stores, we need to implement our own crawlers to dynamically access, crawl, parse and download to obtain the data.
Another challenge is the access rate limits imposed by various app stores. These limitations are intended to prevent server overload but complicate the crawling process. For example, 
Google Play and Steam impose strict access rate controls.
To circumvent these restrictions, we implement a distributed crawling system using multiple authenticated agents 
to parallelize crawling. 
This challenge is mitigated by employing strategies such as spreading requests over extended periods.

\subsubsection{Data Aggregation Across Platforms}
Metadata consistency across platforms is another challenge that arises when aggregating data from multiple sources. While most app stores provide diversified metadata such as application name, developer, and description, there is often variability in the level of detail, the types and labels of attributes. 
Moreover, the metadata of applications that appear on multiple platforms can be inconsistent.
Additional effort is required to duplicate, synthesize and normalize all data across platforms.

%% file: files/free_dataset.tex
\section{\xrappdataset}

\begin{table}[t!]
	\centering
	\caption{The current state of \xrappdataset}
        \vspace{-0.8em}
	\label{table:xrzoo}
	\resizebox{\linewidth}{!}{
	\begin{threeparttable}
	\begin{tabular}{lrrrr}
		\toprule
		\textbf{Marketplace} & \multicolumn{1}{c}{\begin{tabular}[c]{@{}c@{}}\textbf{XR} \\ \textbf{Type(s)}\end{tabular}} & \multicolumn{1}{c}{\begin{tabular}[c]{@{}c@{}}\textbf{App} \\ \textbf{Type(s)\tnote{1}}\end{tabular}} & \multicolumn{1}{c}{\begin{tabular}[c]{@{}c@{}}\textbf{\# of Free}\\ \textbf{Apps}\end{tabular}} & \multicolumn{1}{c}{\begin{tabular}[c]{@{}c@{}}\textbf{\# of Apps}\\ \textbf{Crawled}\end{tabular}} \\
		\cmidrule(){1-5}
		Meta Horizon Store~\cite{website:oculus-app-store} & VR/MR & S, P, M & 861 & 2,153 \\
		VIVEPORT~\cite{website:viveport} & VR & S, P & 709 & 2,712 \\
		SideQuest~\cite{website:sidequest} & VR/MR/AR & S, P, M & 4,783 & 6,246 \\
		Meta Horizon App Lab~\cite{website:oculus-app-lab} & VR/MR & S & 532 & 758 \\
            Vision Pro App Store~\cite{website:vision-pro} & MR & S, P, M & 148 & 311 \\
             & MR supported & S, P, M & 309 & 370 \\
		\cdashlinelr{1-5}
		Steam~\cite{website:steam-app-store-vr} & VR & P & 1,301 & 6,140 \\
             & VR supported & P & 227 & 1,306 \\
            Microsoft Store~\cite{website:hololens} & MR & S, P, M & 353 & 411 \\
		\cdashlinelr{1-5}
		Google Play~\cite{website:google-play} & VR & M & 211 & 443 \\
             & AR & M & 166 & 166 \\
		iOS App Store~\cite{website:apple-app-store} & VR & M & 175 & 201 \\
             & AR & M & 413 & 428 \\
		\cmidrule(){1-5}
            Total & VR/MR/AR & S, P, M & 10,188 & 21,645 \\ \bottomrule
	\end{tabular}
    \begin{tablenotes}
        \item[1] Data collection time: October 2024.
	    \item[1] S, P, and M stand for applications running on standalone, PC-assisted, and mobile device sets, respectively. 
    \end{tablenotes}
    \end{threeparttable}
    }
\end{table}

Table~\ref{table:xrzoo} summarizes the overall statistics of \xrappdataset.
In total, \xrappdataset consists of \xrappdatasetnumber XR applications from \xrzooappstorenumber mainstream app stores, representing a broad range of VR, AR, and MR content. These applications span various XR platforms, including standalone XR devices, PC-assisted XR devices (which rely on an external PC for computation resources), and mobile phone platforms, ensuring the diversity and representativeness of the dataset. This dataset provides a comprehensive foundation for SE research trends in XR application development, testing and bug detection.

\begin{figure*}[t!] 
 	\centering 
 	\includegraphics[width=1.7\columnwidth]{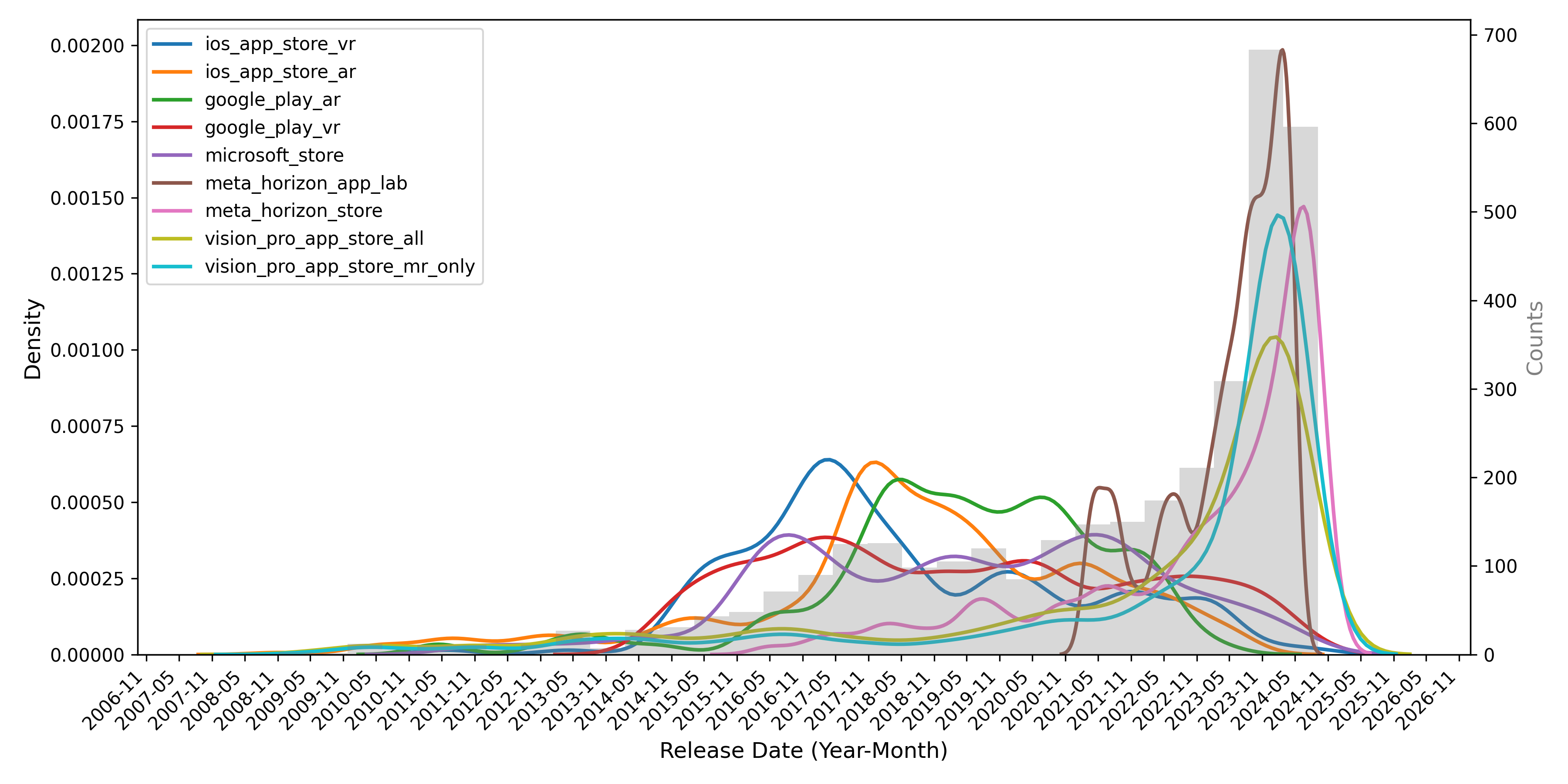} 
        \vspace{-1em}
 	\caption{Density and counts of XR application release dates in \xrappdataset}
 	\label{fig:density_and_histogram}
       \vspace{-1.5em}
 \end{figure*}

Figure~\ref{fig:density_and_histogram} demonstrates the temporal distribution of XR application releases across nine app stores and reveals key trends, with peaks reflecting growth in AR, VR, and MR development. Post-2020, platforms like Meta Horizon App Lab and Vision Pro saw sharp increases, while iOS and Google Play exhibit steady growth, highlighting mobile XR investment. This dataset captures critical shifts in XR development, supporting longitudinal and cross-platform analyses.

\begin{figure*}[t!] 
 	\centering 
 	\includegraphics[width=1.7\columnwidth]{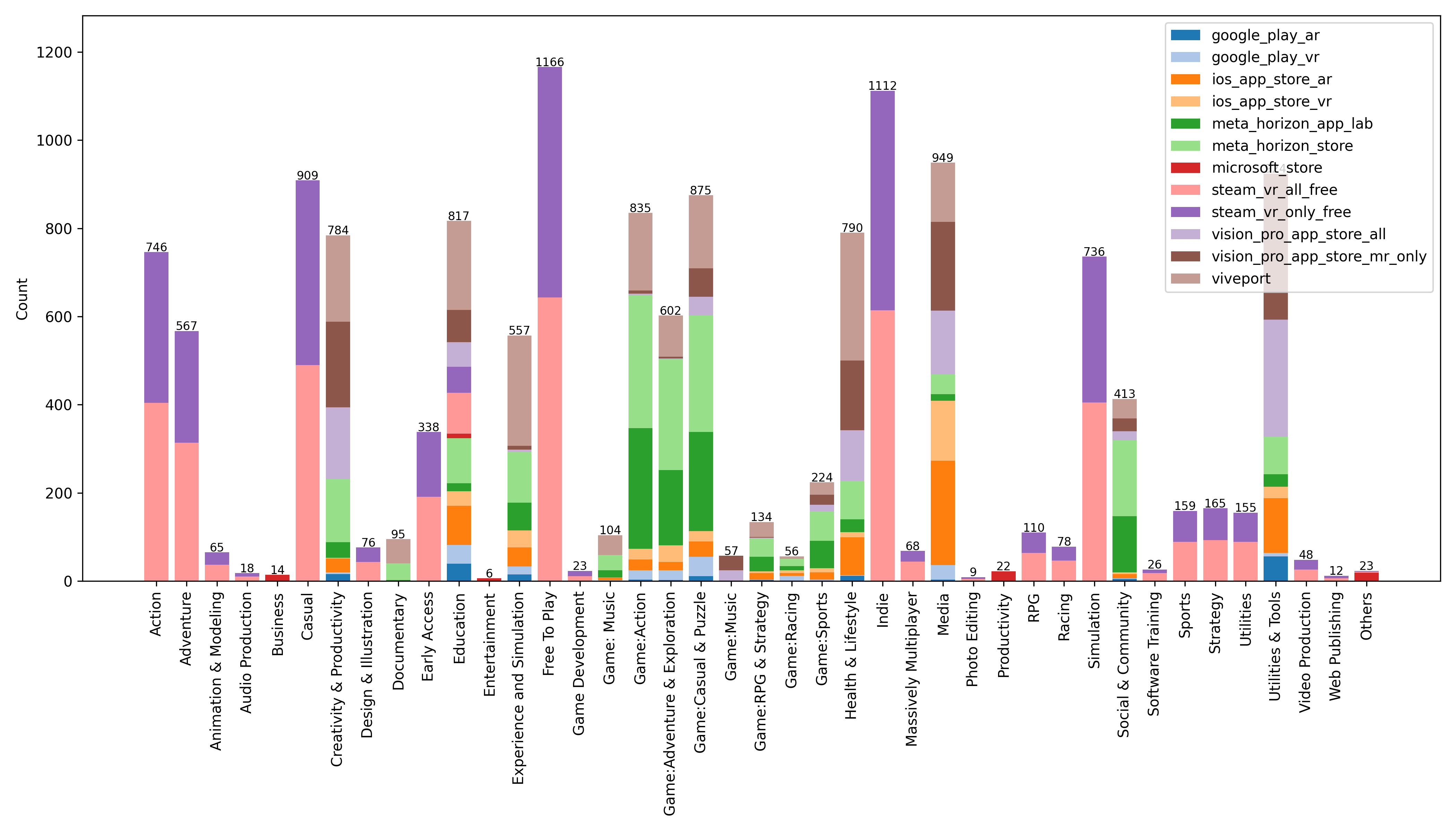} 
        \vspace{-0.8em}
 	\caption{XR application category distribution in \xrappdataset, the category types are aligned among all merketplaces}
 	\label{fig:app_category_distribution}
       \vspace{-1.5em}
 \end{figure*}

Figure~\ref{fig:app_category_distribution} depicts the XR application category distribution in \xrappdataset. It showcases diverse XR use cases, dominated by entertainment categories like Simulation (3,961 apps), Indie, and Action, alongside strong representation in education and productivity. Niche domains such as Animation \& Modeling appear less populated. Steam VR and Vision Pro lead in hosting diverse applications, underscoring \xrappdataset's value for trend and opportunity analysis in XR software.

Table~\ref{table:xrzoo-user-review} shows the user review statistics of applications in \xrappdataset. Vision Pro leads with the highest average (116,921.80) and maximum (17,645,748) reviews, while platforms like Meta Horizon and Viveport report minimal user feedback. iOS App Store reviews show high averages but skewed distributions, suggesting dominance by few apps. Google Play AR and VR present more balanced metrics with moderate medians. These patterns reveal varied user engagement, offering insights for platform-specific XR application analysis.

\begin{table}[t!]
	\centering
	\caption{User review statistics of apps in \xrappdataset}
        \vspace{-0.8em}
	\label{table:xrzoo-user-review}
	\resizebox{\linewidth}{!}{
 \begin{tabular}{lrrrrrr}
\toprule
\textbf{App Source} & \textbf{Mean} & \textbf{Median} & \textbf{std} & \textbf{Max} \\
\midrule
google\_play\_ar & 898.15 & 8 & 6,082.61  & 66,624 \\
google\_play\_vr & 810.15 & 61 & 4,576.59  & 61,802 \\
ios\_app\_store\_ar & 24,732.44 & 163 & 146,414.61  & 1,750,536 \\
ios\_app\_store\_vr & 78,476.50 & 39 & 975,656.04  & 12,891,791 \\
meta\_horizon\_app\_lab & 21.44 & 14 & 18.25  & 50 \\
meta\_horizon\_store & 16.62 & 5 & 20.11  & 50 \\
vision\_pro\_all & 116,921.80 & 10 & 1,017,659.70  & 17,645,748 \\
vision\_pro\_mr\_only & 88,319.44 & 0 & 885,662.26 & 17,645,748 \\
viveport & 9.07 & 3 & 34.93  & 686 \\
\bottomrule
\end{tabular}
    }
\end{table}

\section{Use Cases}

By bridging the gap between the increasing complexity of XR ecosystems and the need for reliable, scalable tools and methods, \xrappdataset serves as an indispensable resource for fostering innovation and progress in XR application development and research.
For researchers, it offers the opportunity to develop and benchmark automated tools for interactive testing, tailored to the spatial computing and real-time interaction complexities inherent to XR environments. It also facilitates advanced studies in bug detection and defect analysis by providing both the application and the metadata needed to identify and address defect patterns across varied XR platforms. Privacy and security auditing is another key area, with the dataset enabling static/dynamic analysis to conduct compliance assessments against privacy policy regulations. 
\xrappdataset can also help XR clone detection research on binary files by providing large-scale application collections.
Additionally, the dataset is a valuable resource for performance optimization research, allowing longitudinal and cross-platform studies to identify and resolve latency and efficiency issues critical to XR experiences. Developers can benefit from \xrappdataset by using it to benchmark application performance, refine debugging workflows, and adopt best practices derived from comparative analyses of applications across platforms like Meta Quest, Vision Pro, and SteamVR. The dataset also supports usability studies and application classification tasks, enabling researchers to develop taxonomies and machine-learning models that capture trends in functionality, user engagement, and design. 
Furthermore, \xrappdataset empowers evolutionary studies, providing upload dates and engagement metrics that allow for investigations into lifecycle patterns, adoption rates, and technological advancements.

%% file: files/conclusion.tex
\section{Conclusion}

In this paper, we presented \xrappdataset, the first large-scale dataset of 12,528 XR applications spanning diverse platforms, categories, and ecosystems. \xrappdataset addresses critical gaps in software engineering research by enabling studies in testing, bug detection, performance analysis, and more, while ensuring comprehensive coverage of the fragmented XR landscape.
Although limitations exist, such as the exclusion of proprietary and paid apps, future work will expand the dataset with richer metadata and longitudinal updates. We believe \xrappdataset will serve as a foundational resource, fostering advancements in the development and maintenance of reliable and efficient XR software.
\xrappdataset is publicly released and actively maintained, which can be accessed via \href{https://sites.google.com/view/xrzoo}{\color{blue}{https://sites.google.com/view/xrzoo}}.

%% file: msr.bbl
\begin{thebibliography}{10}
\providecommand{\url}[1]{#1}
\csname url@samestyle\endcsname
\providecommand{\newblock}{\relax}
\providecommand{\bibinfo}[2]{#2}
\providecommand{\BIBentrySTDinterwordspacing}{\spaceskip=0pt\relax}
\providecommand{\BIBentryALTinterwordstretchfactor}{4}
\providecommand{\BIBentryALTinterwordspacing}{\spaceskip=\fontdimen2\font plus
\BIBentryALTinterwordstretchfactor\fontdimen3\font minus \fontdimen4\font\relax}
\providecommand{\BIBforeignlanguage}[2]{{%
\expandafter\ifx\csname l@#1\endcsname\relax
\typeout{** WARNING: IEEEtran.bst: No hyphenation pattern has been}%
\typeout{** loaded for the language `#1'. Using the pattern for}%
\typeout{** the default language instead.}%
\else
\language=\csname l@#1\endcsname
\fi
#2}}
\providecommand{\BIBdecl}{\relax}
\BIBdecl

\bibitem{bhowmik2024virtual}
A.~K. Bhowmik, ``Virtual and augmented reality: Human sensory-perceptual requirements and trends for immersive spatial computing experiences,'' \emph{Journal of the Society for Information Display}, vol.~32, no.~8, pp. 605--646, 2024.

\bibitem{DBLP:journals/csur/AudaGFMS24}
\BIBentryALTinterwordspacing
J.~Auda, U.~Gruenefeld, S.~Faltaous, S.~Mayer, and S.~Schneegass, ``A scoping survey on cross-reality systems,'' \emph{{ACM} Comput. Surv.}, vol.~56, no.~4, pp. 83:1--83:38, 2024. [Online]. Available: \url{https://doi.org/10.1145/3616536}
\BIBentrySTDinterwordspacing

\bibitem{dionisio20133d}
J.~D.~N. Dionisio, W.~G.~B. Iii, and R.~Gilbert, ``3d virtual worlds and the metaverse: Current status and future possibilities,'' \emph{ACM computing surveys (CSUR)}, vol.~45, no.~3, pp. 1--38, 2013.

\bibitem{report:global-augmented-virtual-reality-market-size-msr}
T.~Alsop, ``Extended reality (xr) market size worldwide from 2021 to 2026,'' \emph{Statista, maart}, 2023.

\bibitem{ziker2021cross}
C.~Ziker, B.~Truman, and H.~Dodds, ``Cross reality (xr): Challenges and opportunities across the spectrum,'' \emph{Innovative learning environments in STEM higher education: Opportunities, challenges, and looking forward}, pp. 55--77, 2021.

\bibitem{li2023towards}
S.~Li, L.~Wei, Y.~Liu, C.~Gao, S.-C. Cheung, and M.~R. Lyu, ``Towards modeling software quality of virtual reality applications from users' perspectives,'' \emph{arXiv preprint arXiv:2308.06783}, 2023.

\bibitem{paper:issre-webxr-empirical}
S.~Li, Y.~Wu, Y.~Liu, D.~Wang, M.~Wen, Y.~Tao, Y.~Sui, and Y.~Liu, ``{An Exploratory Study of Bugs in Extended Reality Applications on the Web},'' in \emph{{ISSRE}}.\hskip 1em plus 0.5em minus 0.4em\relax {IEEE}, 2020, pp. 172--183.

\bibitem{nair2024berkeley}
V.~Nair, W.~Guo, R.~Wang, J.~F. O'Brien, L.~Rosenberg, and D.~Song, ``Berkeley open extended reality recordings 2023 (boxrr-23): 4.7 million motion capture recordings from 105,000 xr users,'' \emph{IEEE Transactions on Visualization and Computer Graphics}, 2024.

\bibitem{wen2024vr}
E.~Wen, C.~Gupta, P.~Sasikumar, M.~Billinghurst, J.~Wilmott, E.~Skow, A.~Dey, and S.~Nanayakkara, ``Vr. net: A real-world dataset for virtual reality motion sickness research,'' \emph{IEEE Transactions on Visualization and Computer Graphics}, 2024.

\bibitem{li2024grounded}
S.~Li, B.~Li, Y.~Liu, C.~Gao, J.~Zhang, S.-C. Cheung, and M.~R. Lyu, ``Grounded gui understanding for vision based spatial intelligent agent: Exemplified by virtual reality apps,'' \emph{arXiv preprint arXiv:2409.10811}, 2024.

\bibitem{qin2024utilizing}
X.~Qin and G.~Weaver, ``Utilizing generative ai for vr exploration testing: A case study,'' in \emph{Proceedings of the 39th IEEE/ACM International Conference on Automated Software Engineering Workshops}, 2024, pp. 228--232.

\bibitem{yang2024towards}
X.~Yang, Y.~Wang, T.~Rafi, D.~Liu, X.~Wang, and X.~Zhang, ``Towards automatic oracle prediction for ar testing: Assessing virtual object placement quality under real-world scenes,'' in \emph{Proceedings of the 33rd ACM SIGSOFT International Symposium on Software Testing and Analysis}, 2024, pp. 717--729.

\bibitem{li2024less}
S.~Li, C.~Gao, J.~Zhang, Y.~Zhang, Y.~Liu, J.~Gu, Y.~Peng, and M.~R. Lyu, ``Less cybersickness, please: Demystifying and detecting stereoscopic visual inconsistencies in virtual reality apps,'' \emph{Proceedings of the ACM on Software Engineering}, vol.~1, no. FSE, pp. 2167--2189, 2024.

\bibitem{guo2024empirical}
H.~Guo, H.-N. Dai, X.~Luo, Z.~Zheng, G.~Xu, and F.~He, ``An empirical study on oculus virtual reality applications: Security and privacy perspectives,'' in \emph{Proceedings of the IEEE/ACM 46th International Conference on Software Engineering}, 2024, pp. 1--13.

\bibitem{trimananda2022ovrseen}
R.~Trimananda, H.~Le, H.~Cui, J.~T. Ho, A.~Shuba, and A.~Markopoulou, ``$\{$OVRseen$\}$: Auditing network traffic and privacy policies in oculus $\{$VR$\}$,'' in \emph{31st USENIX security symposium (USENIX security 22)}, 2022, pp. 3789--3806.

\bibitem{paper:androzoo}
\BIBentryALTinterwordspacing
K.~Allix, T.~F. Bissyand{\'{e}}, J.~Klein, and Y.~L. Traon, ``Androzoo: collecting millions of android apps for the research community,'' in \emph{Proceedings of the 13th International Conference on Mining Software Repositories, {MSR} 2016, Austin, TX, USA, May 14-22, 2016}, M.~Kim, R.~Robbes, and C.~Bird, Eds.\hskip 1em plus 0.5em minus 0.4em\relax {ACM}, 2016, pp. 468--471. [Online]. Available: \url{https://doi.org/10.1145/2901739.2903508}
\BIBentrySTDinterwordspacing

\bibitem{paper:androzooopen}
\BIBentryALTinterwordspacing
P.~Liu, L.~Li, Y.~Zhao, X.~Sun, and J.~Grundy, ``Androzooopen: Collecting large-scale open source android apps for the research community,'' in \emph{{MSR} '20: 17th International Conference on Mining Software Repositories, Seoul, Republic of Korea, 29-30 June, 2020}, S.~Kim, G.~Gousios, S.~Nadi, and J.~Hejderup, Eds.\hskip 1em plus 0.5em minus 0.4em\relax {ACM}, 2020, pp. 548--552. [Online]. Available: \url{https://doi.org/10.1145/3379597.3387503}
\BIBentrySTDinterwordspacing

\bibitem{paper:android-dataset-3}
\BIBentryALTinterwordspacing
D.~E. Krutz, M.~Mirakhorli, S.~A. Malachowsky, A.~Ruiz, J.~Peterson, A.~Filipski, and J.~Smith, ``A dataset of open-source android applications,'' in \emph{12th {IEEE/ACM} Working Conference on Mining Software Repositories, {MSR} 2015, Florence, Italy, May 16-17, 2015}, M.~D. Penta, M.~Pinzger, and R.~Robbes, Eds.\hskip 1em plus 0.5em minus 0.4em\relax {IEEE} Computer Society, 2015, pp. 522--525. [Online]. Available: \url{https://doi.org/10.1109/MSR.2015.79}
\BIBentrySTDinterwordspacing

\bibitem{paper:playmydata}
\BIBentryALTinterwordspacing
A.~D'Angelo, C.~D. Sipio, C.~Politowski, and R.~Rubei, ``Playmydata: a curated dataset of multi-platform video games,'' in \emph{21st {IEEE/ACM} International Conference on Mining Software Repositories, {MSR} 2024, Lisbon, Portugal, April 15-16, 2024}, D.~Spinellis, A.~Bacchelli, and E.~Constantinou, Eds.\hskip 1em plus 0.5em minus 0.4em\relax {ACM}, 2024, pp. 525--529. [Online]. Available: \url{https://doi.org/10.1145/3643991.3644869}
\BIBentrySTDinterwordspacing

\bibitem{paper:conf/msr/dAragonaBLSAABBKLTWNQRAMCT24}
\BIBentryALTinterwordspacing
D.~A. d'Aragona, A.~Bakhtin, X.~Li, R.~Su, L.~Adams, E.~Aponte, F.~Boyle, P.~Boyle, R.~Koerner, J.~Lee, F.~Tian, Y.~Wang, J.~Nyyss{\"{o}}l{\"{a}}, E.~Quevedo, M.~S. Rahaman, A.~S. Abdelfattah, M.~M{\"{a}}ntyl{\"{a}}, T.~Cern{\'{y}}, and D.~Taibi, ``A dataset of microservices-based open-source projects,'' in \emph{21st {IEEE/ACM} International Conference on Mining Software Repositories, {MSR} 2024, Lisbon, Portugal, April 15-16, 2024}, D.~Spinellis, A.~Bacchelli, and E.~Constantinou, Eds.\hskip 1em plus 0.5em minus 0.4em\relax {ACM}, 2024, pp. 504--509. [Online]. Available: \url{https://doi.org/10.1145/3643991.3644890}
\BIBentrySTDinterwordspacing

\bibitem{paper:libscout}
M.~Backes, S.~Bugiel, and E.~Derr, ``{Reliable Third-Party Library Detection in Android and its Security Applications},'' in \emph{{CCS}}.\hskip 1em plus 0.5em minus 0.4em\relax {ACM}, 2016, pp. 356--367.

\bibitem{website:oculus-app-store}
``{Meta Horizon Store},'' \url{https://www.meta.com/experiences/}, 2024.

\bibitem{website:viveport}
``{VIVEPORT},'' \url{https://www.viveport.com/}, 2024.

\bibitem{website:sidequest}
``{SideQuest},'' \url{https://sidequestvr.com/}, 2024.

\bibitem{website:oculus-app-lab}
``{Meta Horizon App Lab},'' \url{https://developer.oculus.com/blog/introducing-app-lab-a-new-way-to-distribute-oculus-quest-apps/}, 2024.

\bibitem{website:vision-pro}
``{Vision Pro},'' \url{https://apps.apple.com/us/vision/}, 2024.

\bibitem{website:steam-app-store-vr}
``{VR Content on Steam App Store},'' \url{https://store.steampowered.com/search/?vrsupport=401}, 2024.

\bibitem{website:hololens}
``{HoloLens},'' \url{https://www.microsoft.com/en-us/store/collections/hlgettingstarted/hololens}, 2024.

\bibitem{website:google-play}
``{Google Play},'' \url{https://play.google.com/}, 2024.

\bibitem{website:apple-app-store}
``{iOS App Store},'' \url{https://www.apple.com/app-store/}, 2024.

\end{thebibliography}
